# Antiferromagnetism in Ni-Based Superconductors

Xiaorong Zhou[1,#], Xiaowei Zhang[2,#], Jiabao Yi[3,#], Peixin Qin[1,#], Zexin Feng[1,#], Peiheng Jiang[4], Zhicheng Zhong[4], Han Yan[1], Xiaoning Wang[1], Hongyu Chen[1], Haojiang Wu[1], Xin Zhang[1], Ziang Meng[1], Xiaojiang Yu[5], Mark B. H. Breese[5,6], Jiefeng Cao[7], Jingmin Wang[1], Chengbao Jiang[1], Zhiqi Liu[1,*]

1. School of Materials Science and Engineering, Beihang University, Beijing 100191, China
2. School of Electrical Engineering and Computer Science, Ningbo University, Ningbo 315211, China
3. Global Innovative Centre for Advanced Nanomaterials School of Engineering, The University of Newcastle, Callaghan, NSW 2308, Australia
4. Key Laboratory of Magnetic Materials and Devices and Zhejiang Province Key Laboratory of Magnetic Materials and Application Technology, Ningbo Institute of Materials Technology and Engineering (NIMTE), Chinese Academy of Sciences, Ningbo 315201, China
5. Singapore Synchrotron Light Source, National University of Singapore, 5 Research Link, Singapore 117603, Singapore
6. Department of Physics, National University of Singapore, Singapore 117542, Singapore
7. Shanghai Synchrotron Radiation Facility, Shanghai Advanced Research Institute, Chinese Academy of Sciences, Shanghai 201204, China

[#]These authors contributed equally to this work.
*email: zhiqi@buaa.edu.cn

**Keywords:** Ni-based superconductors; nickelates; magnetic order; antiferromagnetism; exchange bias

**Due to the lack of any magnetic order down to 1.7 K in the parent bulk compound $NdNiO_2$, the recently discovered 9-15 K superconductivity in the infinite-layer $Nd_{0.8}Sr_{0.2}NiO_2$ thin films has provided an exciting playground for unearthing new superconductivity mechanisms. In this letter, we report the successful synthesis of a series of superconducting $Nd_{0.8}Sr_{0.2}NiO_2$ thin films ranging from 8 to 40 nm. We observe the large exchange bias effect between the superconducting $Nd_{0.8}Sr_{0.2}NiO_2$ films and a thin ferromagnetic layer, which suggests the existence of the antiferromagnetic order. Furthermore, the existence of the antiferromagnetic order is evidenced by X-ray magnetic linear dichroism measurements. These experimental results are fundamentally critical for the current field.**

## 1. Introduction

Unlike the well-known high-temperature superconductors such as cuprates and Fe-based superconductors that were first discovered in bulk compounds[1,2], the recently reported superconductivity in an infinite-layer Ni compound is based on nanoscale $Nd_{0.8}Sr_{0.2}NiO_2$ (NSNO) thin films[3], which inevitably involve surface/interface effects from substrates and could be largely different from bulk materials. For example, the newly synthesized NSNO bulk compounds are found to lack in superconductivity[4].

One of the main reasons for the immediate surge of research on NSNO-related superconductivity is that the bulk material of its parent compound $NdNiO_2$ does not exhibit any magnetic order down to 1.7 K[5]. Thus, the superconductivity in NSNO thin films could be a new type of superconducting phenomenon in contrast to high-temperature cuprate superconductors for which the antiferromagnetic order in their parent compounds could be crucial for the formation of Cooper pairs and the electronic structure is different from nickelates[6].

Compared with the stable perovskite $NdNiO_3$ phase, the infinite-layer $NdNiO_2$ phase could hardly be stabilized. Instead, a long-period special chemical reduction process is needed to convert $NdNiO_3$ into $NdNiO_2$[5], during which reductive agents such as sodium hydride need to be used. The recent fabrication of the infinite-layer NSNO films was achieved by a similar fashion[3], *i.e.*, chemical reduction of perovskite $Nd_{0.8}Sr_{0.2}NiO_3$ films, which were deposited by pulsed laser deposition on $SrTiO_3$ (STO) substrates at a high oxygen pressure of 150 mTorr, to NSNO films by reductive agent $CaH_2$.

Moreover, the chemical reduction was carried out in a Pyrex glass tube at 260-280 ºC for 4-6 h to ensure a sufficient gas-phase reaction. Due to this essential and sophisticated reduction process, the experimental repetition of the superconductivity has been challenging as most of experimental oxide electronic groups may have limited experience in dealing with chemical reduction, for which the amount of the $CaH_2$, the temperature and the reduction time are all important to be optimized. As a result, there have been only few reports on the experimental realization of superconductivity for NSNO films despite of a large number of theoretical studies. Most of them were from the same group that discovered this phenomenon[3,7-10], one is from Gu *et al.*[11] which studied the nature of the superconducting gaps via scanning tunneling microscopy spectroscopy and another one is from Zeng *et al.*[12] which reported the successful fabrication of superconducting NSNO films by reducing $Nd_{0.8}Sr_{0.2}NiO_3$ thin films with $CaH_2$ powder in a vacuum chamber at 340-360 ºC for 80-120 mins. The approach reported in Reference 12 could be more feasible for most of oxide thin film groups as it is conveniently performed in a pulsed laser deposition vacuum chamber with a much shorter time.

## 2. Results and Discussion

Based on our previous experimental study[13], we have further utilized $CaH_2$ powder to chemically reduce 20-nm-thick $Nd_{0.8}Sr_{0.2}NiO_3$ films, which were pre-fabricated to epitaxially grow on STO substrates at an oxygen pressure of 150 mTorr and 700 ºC by pulsed laser deposition, in the pulsed laser deposition vacuum chamber at 350 ºC for 2 h. **Figure 1**a shows a typical X-ray diffraction pattern of an NSNO/STO heterostructure after chemical reduction. The (001) peak located at 26.5º and the (002) peak at 54.6º correspond to an out-of-plane lattice constant $c \sim 3.36$ Å, which is in good agreement with previous experimental reports on the superconducting NSNO thin films[3,12]. The 360º phi scans around the STO (101) and NSNO (101) peaks confirm the epitaxy of the NSNO thin film on the STO substrate (Figure 1b). X-ray reflectivity spectrum shown in Figure 1c reveals that the thin film is smooth.

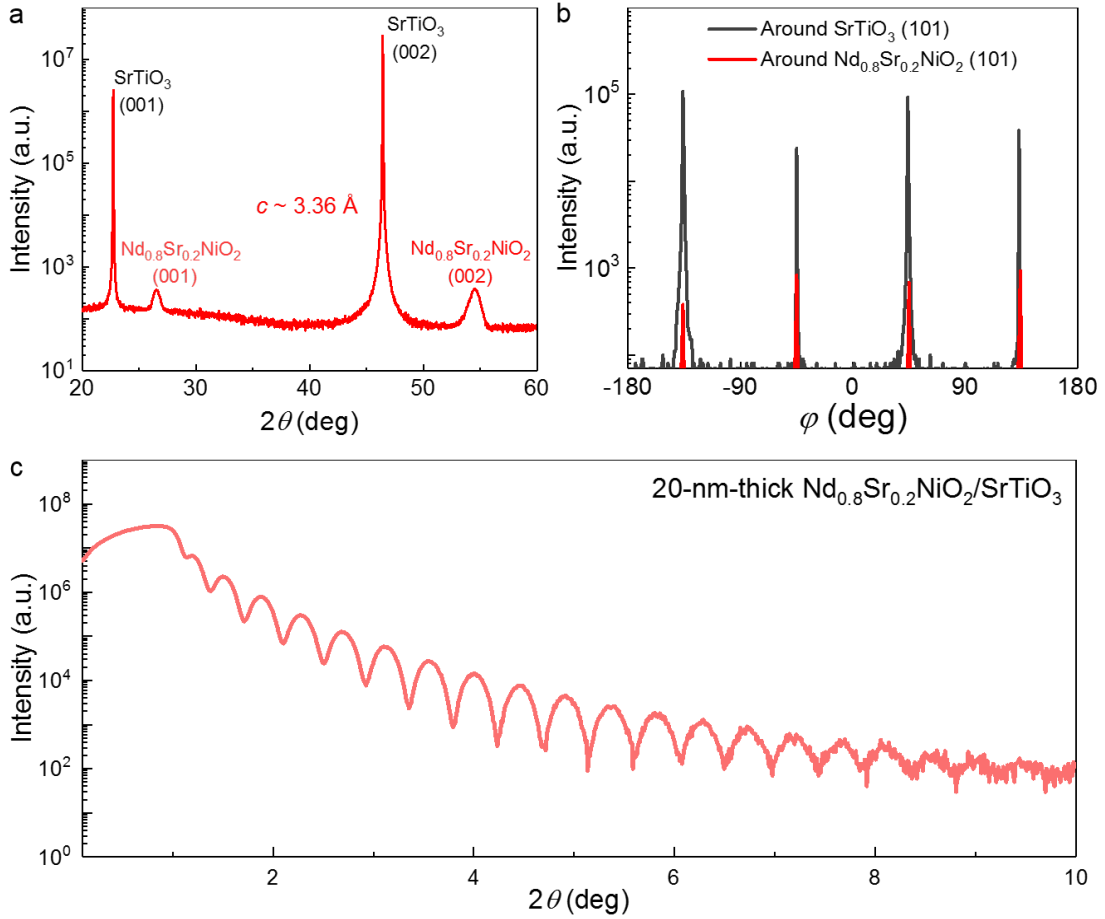

**Figure 1.** Structure of a 20-nm-thick $Nd_{0.8}Sr_{0.2}NiO_2/SrTiO_3$ (NSNO/STO) heterostructure. a) $\theta$-$2\theta$ X-ray diffraction pattern. b) 360º phi scans around the NSNO and STO (101) peaks. c) X-ray reflectivity of the 20 nm NSNO film up to 10º.

**Figure 2**a plots the temperature-dependent resistivity of the NSNO/STO heterostructure. The room-temperature resistivity is ~1.5 mΩ·cm, which is much lower than that of the NSNO thin films reported in Reference 12 but rather comparable with that reported in Reference 3. Above ~150 K, the resistivity is linear with temperature, reminiscence of the temperature-dependent resistivity of a high-temperature superconducting $YBa_2Cu_3O_7$ thin film[14]. The linear temperature dependence is deviated at low temperatures and a resistivity plateau sets on at ~50 K. Finally, the resistivity plateau turns into a superconducting phase transition from $T_{c\ Onset}$ ~ 13.4 K (point of maximum curvature) and the zero-resistance state occurs at $T_{c\ Zero}$ ~ 8.2 K. The superconducting transition temperatures are in good agreement with the previously reported values[3,12].

The temperature-dependent normal-state Hall coefficient $R_H$ obtained from Hall measurements is shown in Figure 2b. $R_H$ is negative for high temperatures but gradually transits to be positive blow ~50 K. This temperature evolution of the Hall coefficient is well akin to what has been observed in Reference 3 & 12.

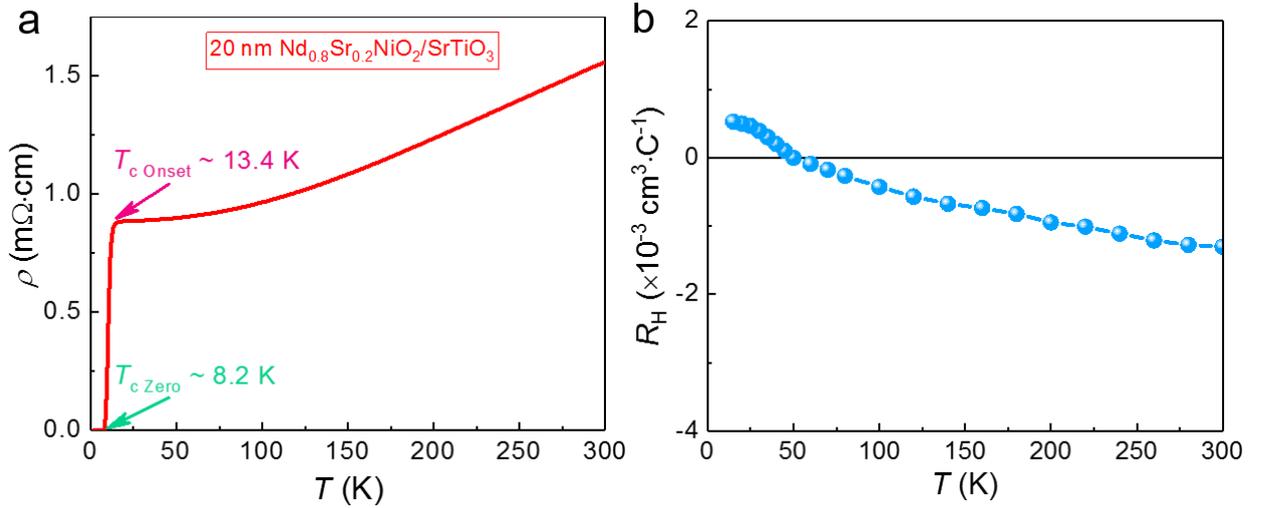

**Figure 2.** Electrical transport properties. a) Temperature-dependent resistivity of the 20-nm-thick NSNO/STO heterostructure. b) Temperature-dependent normal-state Hall coefficient of the NSNO/STO heterostructure.

The magnetic measurements on the NSNO/STO heterostructure reveal predominant diamagnetic signals from STO substrates for all the temperatures ranging from 300 to 2 K, suggesting a negligible magnetic moment of the NSNO film (**Figure S1** & **S2**). Afterwards, we deposited a 5-nm-thick soft ferromagnetic $Co_{90}Fe_{10}$ layer on top of the NSNO/STO heterostructure and then capped the whole stack with a 2-nm-thick Pt layer (schematized in **Figure 3**a) to prevent oxidation of the $Co_{90}Fe_{10}$. The $Co_{90}Fe_{10}$ and Pt metal layers were fabricated by sputtering at room temperature. To induce an artificial easy axis in the $Co_{90}Fe_{10}$, an in-plane field of ~20 mT was applied onto the NSNO/STO heterostructure via NdFeB permanent magnets during the deposition of the $Co_{90}Fe_{10}$ thin layer[15-20]. The clean interface (Figure 3b) between $Co_{90}Fe_{10}$ and NSNO rules out the possible NiO precipitation at the interface. As shown in Figure 3c, low-temperature magnetic measurements on the Pt (2 nm)/$Co_{90}Fe_{10}$ (5 nm)/NSNO (20 nm)/STO stack along the growth field direction of $Co_{90}Fe_{10}$ demonstrates the exchange bias effect, which can be reversed by a negative cooling field (**Figure S3**).

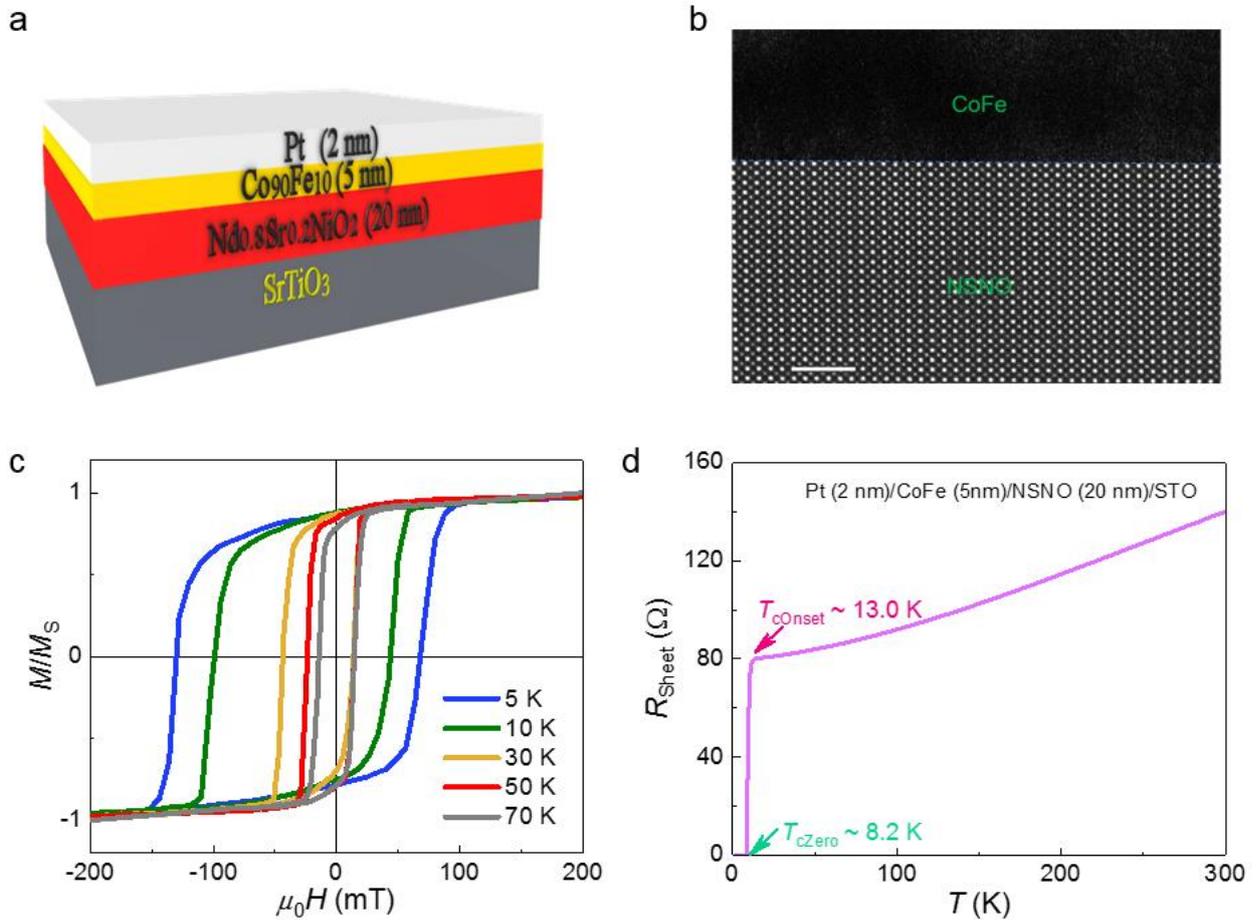

**Figure 3.** Exchange bias. a) Schematic of a Pt (2 nm)/Co$_{90}$Fe$_{10}$ (5 nm)/NSNO (20 nm)/STO heterostructure. b) Cross-section transmission electron microscopy of a Co$_{90}$Fe$_{10}$/NSNO interface. The scale bar stands for 2 nm. c) Normalized magnetization versus magnetic field at different temperatures obtained by field cooling measurements with an in-plane cooling field of +500 mT and the cooling process was started from 300 K. The measurement magnetic field was applied in plane as well. d) Temperature-dependent sheet resistance for the Pt (2 nm)/Co$_{90}$Fe$_{10}$ (5 nm)/NSNO (20 nm)/STO heterostructure.

In addition to the well-established existence of the exchange bias effect for antiferromagnet/ferromagnet interfaces, the exchange bias could also exist for soft ferromagnetic/hard ferromagnet[21], ferromagnet/ferrimagnet interface systems[22] and ferromagnet/spin glass interface systems for which the blocking temperatures are quite close to the spin freezing temperatures[23]. To examine the possible net magnetic moment in the NSNO layer, X-ray magnetic circular dichroism (XMCD) measurements were performed at 6 K and higher temperatures up to 300 K on the Ni L edges for a Pt (2 nm)/Co$_{90}$Fe$_{10}$ (5 nm)/NSNO (20

nm)/STO heterostructure. The Ni L edges were detected in the total electron yield (TEY) mode with a grazing incidence and in-plane magnetic field of 1 T. The absence of any XMCD signal (**Figure S4** & **S5**) over a large temperature range ruled out the possibility of a net ferro/ferri-magnetic moment or a spin glass state in the NSNO layer induced by $Co_{90}Fe_{10}$ deposition. Therefore, these results imply the antiferromagnetic order in the NSNO thin film. The exchange bias field is ~5 mT at 50 K and largely increases to ~32 mT at 5 K. In addition, below the blocking temperature, the coercivity field of $Co_{90}Fe_{10}$ is remarkably enhanced (**Figure S6**) in the Pt (2 nm)/$Co_{90}Fe_{10}$ (5 nm)/NSNO (20 nm)/STO heterostructure compared with the reference Pt (2 nm)/$Co_{90}Fe_{10}$ (5 nm)/STO sample despite of the same saturation magnetization (**Figure S7**) of $Co_{90}Fe_{10}$ for the two types of heterostructures. After the deposition of the ferromagnetic $Co_{90}Fe_{10}$ layer, the superconducting behavior of the NSNO film does not obviously change (Figure 3d), which illustrates the robust nature of the NSNO superconductivity over interfacial magnetic perturbations.

Interestingly, the exchange bias effect exists below the superconducting transition temperature of the NSNO film as well, hence demonstrating the coexistence of the antiferromagnetism and superconductivity in an NSNO thin film. Further detailed magnetic measurements shown in **Figure 4**a reveal a blocking temperature, above which the exchange bias effect disappears, of ~65 K for the NSNO/$Co_{90}Fe_{10}$ bilayer system. In contrast, the reference samples by depositing $Co_{90}Fe_{10}$ onto STO substrates or non-superconducting $Nd_{0.8}Sr_{0.2}NiO_3$ films exhibit zero exchange bias. In addition, the chemical reduction process for $Nd_{0.8}Sr_{0.2}NiO_3$ films to achieve uniform NSNO films seems not optimal to reach a smooth surface of parent-phase $NdNiO_2$ film, which exhibit large surface roughness and consequently zero exchange bias with $Co_{90}Fe_{10}$.

As Co and Fe are chemically reactive materials, they could be easily oxidized if the deposition process has residual oxygen or the interface between $Co_{90}Fe_{10}$ and NSNO becomes active, thus leading to the degradation of magnetic moments and exchange bias. To investigate these aspects, we first measured the magnetic moment of the $Co_{90}Fe_{10}$ film in an exchange-based Pt (2 nm)/$Co_{90}Fe_{10}$ (5 nm)/NSNO (20 nm)/STO heterostructure and found that the room-temperature saturation magnetization of ~1310 emu/cc (**Figure S7**) is rather comparable with the previously reported intrinsic magnetization 1302 emu/cc of $Co_{90}Fe_{10}$[24]. The magnetization loops of the reference Pt (2 nm)/$Co_{90}Fe_{10}$ (5 nm)/STO sample for low temperatures are shown in **Figure S8**. Secondly, to directly probe any possible oxidation states of Co and Fe in the $Co_{90}Fe_{10}$ films of our Pt (2 nm)/$Co_{90}Fe_{10}$ (5 nm)/NSNO (20 nm)/STO samples, we performed X-ray absorption spectroscopy (XAS) measurements with the TEY mode with a normal incidence and a circularly polarized light. Meanwhile, a Pt (2 nm)/$Co_{90}Fe_{10}$ (5 nm)/STO sample that does not exhibit any exchange bias effect was measured to serve as a reference for the $Co_{90}Fe_{10}$/NSNO interface. As shown in Figure 4b & 4c, room-temperature XAS spectra of Co and Fe L edges exhibit clean metallic features and no signature of any oxidation peak is seen for both types of heterostructures. In addition, the field-dependent Co $L_3$ edge XMCD signals collected by a grazing incidence exhibits a similar loop shift at 6 K (**Figure S9**). This indicates that $Co_{90}Fe_{10}$ is not oxidized in our samples and suggests that the $Co_{90}Fe_{10}$/STO interface and $Co_{90}Fe_{10}$/NSNO interface is similar in terms of chemical stability of $Co_{90}Fe_{10}$. To directly visualize the interface situation between $Co_{90}Fe_{10}$ and NSNO, we prepared a cross-section sample via mechanical polishing and ion milling and then performed high-resolution scanning transmission electron microscopy measurements.

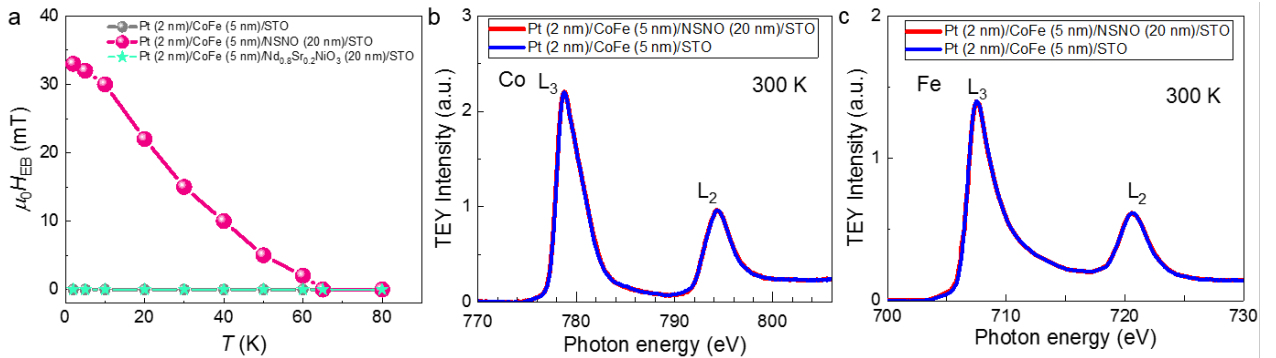

**Figure 4.** Blocking temperature. a) Temperature-dependent exchange bias field of the Pt (2 nm)/$Co_{90}Fe_{10}$ (5 nm)/NSNO (20 nm)/STO heterostructure. The reference samples with a Pt (2 nm)/$Co_{90}Fe_{10}$ (5 nm)/ /STO stacking structure and a Pt (2 nm)/$Co_{90}Fe_{10}$ (5 nm) /$Nd_{0.8}Sr_{0.2}NiO_3$ (20 nm)/ /STO stacking structure exhibit zero exchange

bias. b) X-ray absorption spectra of Co L edges for two different samples. c) X-ray absorption spectra of Fe L edges for two different samples. The spectra were collected by a circularly polarized light in the total electron yield mode with a normal incidence.

To further explore the possible magnetic order of the NSNO thin films, X-ray magnetic linear dichroism (XMLD) measurements were performed. X-ray absorption spectra of Ni $L_{2,3}$ edges for linearly polarized light oriented along different directions relative to the sample surface are shown for different temperatures in **Figure 5**. In solid-state systems, the peak splitting of the L absorption edge is typically related to the hybridization with oxygen[25]. The general shape of the Ni L edges absorption spectra are largely distinct from that of the $Ni^{3+}$ state in $NdNiO_3$[26], which instead is consistent with that of the recent Ni L-edge soft X-ray spectroscopy study on superconducting NSNO films[27]. The obvious difference between the two types of spectra demonstrates a large linear dichroic signal (**Figure S10**). In addition, temperature-dependent $L_3$ peak dichroism displays a critical temperature of ~70 K (**Figure S11**), which is in good agreement with the exchange bias experiment as shown in Figure 4. In addition, the thickness dependent $L_3$ peak dichroism at 6 K is shown in **Figure S12**. Overall, these results confirm the antiferromagnetic ordering in the NSNO thin film.

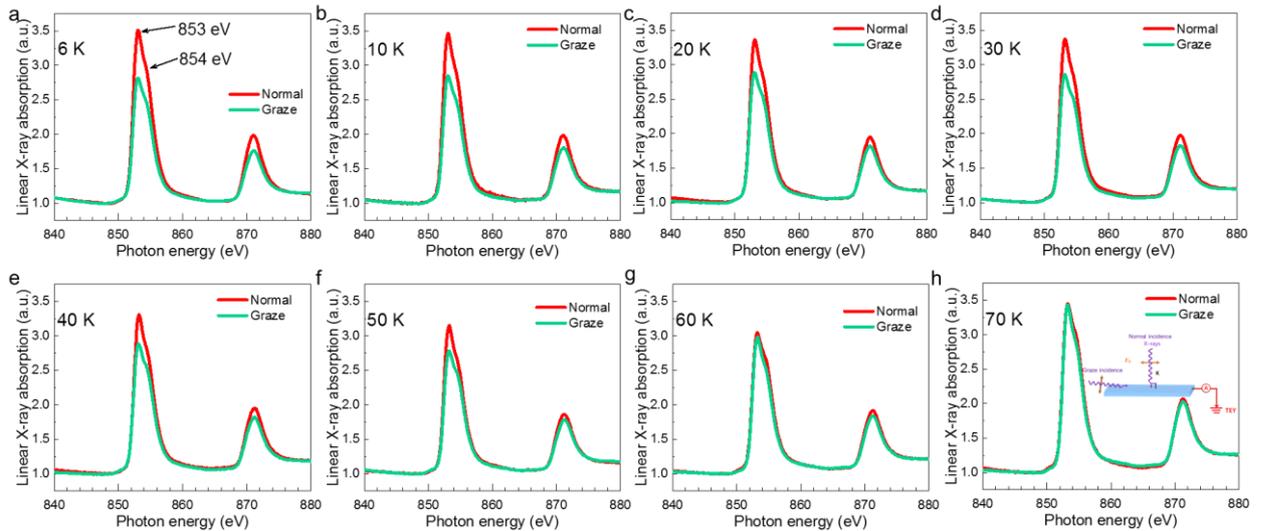

**Figure 5.** X-ray magnetic linear dichroism. a-h) Linear X-ray absorption spectra of Ni $L_{2,3}$ edges in a 20-nm-thick NSNO/STO sample at different temperatures. The red curves correspond to the linearly polarized light oriented perpendicular to the sample surface while the green curves belong to the linearly polarized light oriented almost

parallel to the sample surface with an incidence angle of 30°. Inset in (h): Schematic of the measurement geometry. The pre-edge of all the XAS spectra are normalized to be close to one.

Subsequently, we fabricated a series of NSNO thin films with different thickness ranging from 8 to 40 nm by the same procedure. It was found that $T_{c\ Onset}$ varies from 10.2 to 14.7 K and $T_{c\ Zero}$ alters from 7.7 to 12.7 K (**Figure 6**a). All these films were similarly coated with 5-nm-thick $Co_{90}Fe_{10}$ and 2-nm-thick Pt capping layers for investigating the exchange bias effect. As a result, both the exchange bias field and the blocking temperature exhibit a critical thickness of ~20 nm (Figure 6b & 6c), below which the exchange bias parameter abruptly decreases with reducing thickness. Such behavior has been well established for exchange biased systems and could be related to the thickness-dependent antiferromagnetic domain size and the magnetocrystalline anisotropy energy[28,29]. The slight decrease of the exchange bias field above 20 nm is likely induced by the increased surface roughness of the NSNO films as revealed by X-ray reflectivity measurements. More importantly, the critical temperatures for the existence of linear dichroism are in good agreement with the blocking temperatures (Figure 6c).

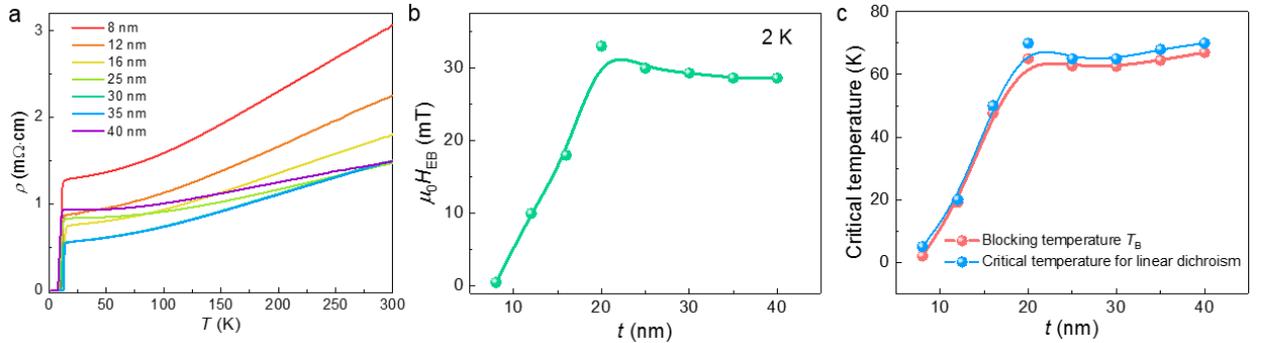

**Figure 6.** Thickness dependence. a) Temperature-dependent resistivity for NSNO thin films with different thickness. b) NSNO-thickness-dependent exchange bias field for the $Co_{90}Fe_{10}$/NSNO system at 2 K. c) NSNO-thickness-dependent blocking temperature for the $Co_{90}Fe_{10}$/NSNO system and NSNO-thickness-dependent critical temperature for the existence of the linear dichroism for the NSNO/STO heterostructures.

The magnetic moment measurements up to 9 T give rise to an estimated field-induced net moment of ~0.01 $\mu_B$/Ni (assuming that Ni atoms carrier all the magnetic moment for the NSNO film) for a 20-nm-thick NSNO film (**Figure S13**). Using a simple spin canting model based on a two-sublattice collinear antiferromagnet picture, we estimated that the upper limit for the Ni local moment is ~0.22 $\mu_B$ (**Note S1**). Although the long-range antiferromagnetic order typically competes with the superconducting ground state such as in cuprates due to the competition of the kinetic energy of free carriers with the exchange interaction, the coexistence of them has been reported for $(Li_{0.8}Fe_{0.2})OHFeSe$[30]. In addition, the strong exchange interaction in nickelates similar to curprates has been recently revealed by Nomura *et al*. (~100 meV for undoped $NdNiO_2$)[31] and Wan *et al.* (82 meV for hole-doped $Nd_{1-x}Sr_xNiO_2$)[32]. Other features of magnetic order in relevant nickelate samples have been also implied[33-35]. It is worth pointing out that both our exchange bias and XMLD experiments are sensitive to the surfaces of NSNO films and it is possible that the antiferromagnetism could only exist on the surface region of superconducting NSNO films. Furthermore, we performed the cross-section conduction mapping[36] for a 35-nm-thick NSNO/STO heterostructure that exhibits the highest $T$c $_{Zero}$ (~12.7 K) via a low-temperature conducting force microscopy setup at 10 K. The uniform cross-section conduction (**Figure S14**) of the NSNO film may shed light on the possibility spatial coexistence of the antiferromagnetism and the superconductivity.

Importantly, it is worth emphasizing that the evidence of an AF order in the bulk $Nd_{1-x}Sr_xNiO_2$ has been experimentally revealed[34] although the sample is non-superconductive. Therefore, the AF order may not be a unique feature for superconducting thin films. On the other hand, if the antiferromagnetic order or antiferromagnetic fluctuations are prerequisites for the emergence of superconductivity, the experimental progress on the magnetic order in bulk $Nd_{1-x}Sr_xNiO_2$ materials could imply that the superconducvity may, in principle, be able to realized in bulk materials as well in future.

In bulk $NdNiO_2$, no magnetic ordering is present down to 1.7 K[5]. Meanwhile, the superconductivity is absent in bulk $Nd_{0.8}Sr_{0.2}NiO_2$ compounds[4,37]. Following this line of possible connection, the antiferromagnetic exchange interaction may be pivotal for the superconductivity in the NSNO thin films. Although the exact origin of the antiferromagnetism in superconducting NSNO thin films needs more detailed studies, for example, our preliminary theoretical calculations found that the surface of NSNO films could help stabilize the antiferromagnetic order (**Note S2**), or interface-induced electronic reconstruction[38-41] could occur at the interface between STO substrates and NSNO thin films similar to the $LaMnO_3$/STO[42,43] and $LaAlO_3$/STO interfaces[44-46], we believe that the experimental observation of the antiferromagnetic order in superconducting NSNO thin films itself is fundamentally crucial so that this work may steer the center of focus for the rapidly developing field on the novel superconductivity in nickelates. In addition to the great interest in the community of condensed matter physics, the successful synthesis of this new type of Ni-based superconductors may pave the way to other important novel material applications such as energy, eletromagnetic and spintronic materials and devices[47-51].

**Experimental Section**

*Sample preparation:* $Nd_{0.8}Sr_{0.2}NiO_3$ thin films were first deposited from a polycrystalline $Nd_{0.8}Sr_{0.2}NiO_3$ ceramic target on (001)-oriented $SrTiO_3$ (STO) single-crystal substrates ($2.5\times5\times0.5$ mm$^3$) in a pulsed laser deposition system (Shenyang Baijujie Scientific Instrument Co., Ltd) with a base pressure of $3\times10^{-8}$ Torr at 700 ºC and 150 mTorr oxygen partial pressure. Afterwards, thin-film samples were loosely contained in Al foils with each piece mixed with 0.1 g $CaH_2$ powder and then moved to the pulsed laser deposition vacuum chamber. The wrapped samples were heated to 350 ºC at 10 ºC/min. During this process, the vacuum chamber was sealed without connecting to the turbo pump and its internal pressure changed from $1.2\times10^{-6}$ to $1.3\times10^{-1}$ Torr during the heating process and finally became stable at $1.5\times10^{-1}$ Torr for the later reduction process. The chemical reduction process was kept at 350 ºC for 2 h. Afterwards, the samples were cooled down to room temperate at 10 ºC/min.

For the exchange bias experiments, the depositions of $Co_{90}Fe_{10}$ and Pt thin films were performed at room temperature in a sputtering system with a based pressure of $7.5\times10^{-9}$ Torr.

The sputtering power for $Co_{90}Fe_{10}$ was 90 W and the Ar pressure was 3 mTorr. The deposition rate was 0.29 Å/s. The sputtering power for Pt was 30 W and the Ar pressure was 3 mTorr. The resulting deposition rate of Pt was 0.28 Å/s.

*X-ray diffraction:* X-ray diffraction patterns were collected by a four-circle PANalytical X-ray diffractometer with a Cu-*Ka*1 wavelength of 1.540598 Å.

*Electrical measurements:* Electrical contacts were made by an ultrasonic wire bonder system with aluminum wires. The linear four-probe electrical resistance measurements were conducted with a 100 µA current in a Quantum Design physical property measurement system.

*Magnetic measurements:* Magnetic moment measurements were carried out by a Quantum Design superconducting quantum interference device magnetometer with the field applied in the thin-film sample plane.

*X-ray magnetic linear dichroism:* X-ray absorption spectra on Ni $L_{2,3}$ edges were collected by linearly polarized light at Singapore Synchrotron Light Source in the total electron yield mode, which was oriented either perpendicular to the sample surface or nearly parallel to the sample surface with an incidence angle of 30°. The difference between the two measurement geometries was normalized to represent the dichroism signal.

*X-ray magnetic circular dichroism:* X-ray absorption spectra on Ni $L_{2,3}$ edges and Co $L_{2,3}$ edges were collected by differently circularly polarized light with a grazing incidence angle of 30° in the total electron yield mode. The difference between the two kinds of absorption spectra was normalized to represent the dichroism signal.

*Transmission electron microscopy:* The cross-section samples were prepared by mechanical polishing and ion milling. The images were taken by a JEM ARM200F setup operated at 200 kV.

**Acknowledgements**
Z.L. acknowledges financial support from the National Natural Science Foundation of China (Nos. 52121001, 51822101, 51861135104 & 51771009). Xiaowei Z. acknowledges financial support from the National Natural Science Foundation of China (Nos. 61974078 & 61704094). J.Y. acknowledges the support of Future Fellowship (Australia Research Council, FT160100205). The authors would like to acknowledge the Singapore Synchrotron Light Source (SSLS) for providing the facility necessary for conducting the research. The Laboratory is a National Research Infrastructure under the National Research Foundation Singapore.

# Supporting Information

## Antiferromagnetism in Ni-Based Superconductors

*Xiaorong Zhou, Xiaowei Zhang, Jiabao Yi, Peixin Qin, Zexin Feng, Peiheng Jiang, Zhicheng Zhong, Han Yan, Xiaoning Wang, Hongyu Chen, Haojiang Wu, Xin Zhang, Ziang Meng, Xiaojiang Yu, Mark B. H. Breese, Jiefeng Cao, Jingmin Wang, Chengbao Jiang, and Zhiqi Liu\**

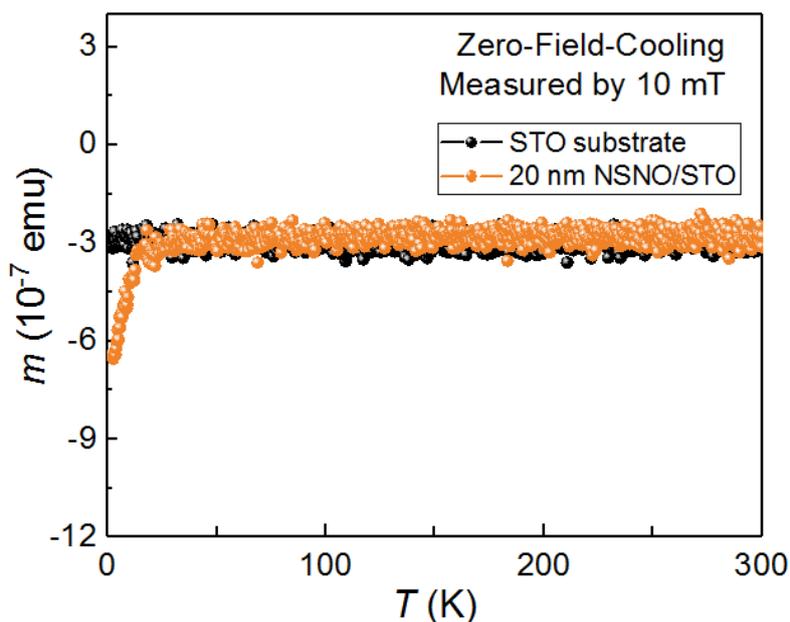

**Figure S1.** Temperature-dependent total magnetic moment of an STO substrate and a 20-nm-thick NSNO/STO heterostructure ($5 \times 2.5 \times 0.5$ mm$^3$) measured by an in-plane field of 10 mT under the zero-field-cooling measurement mode.

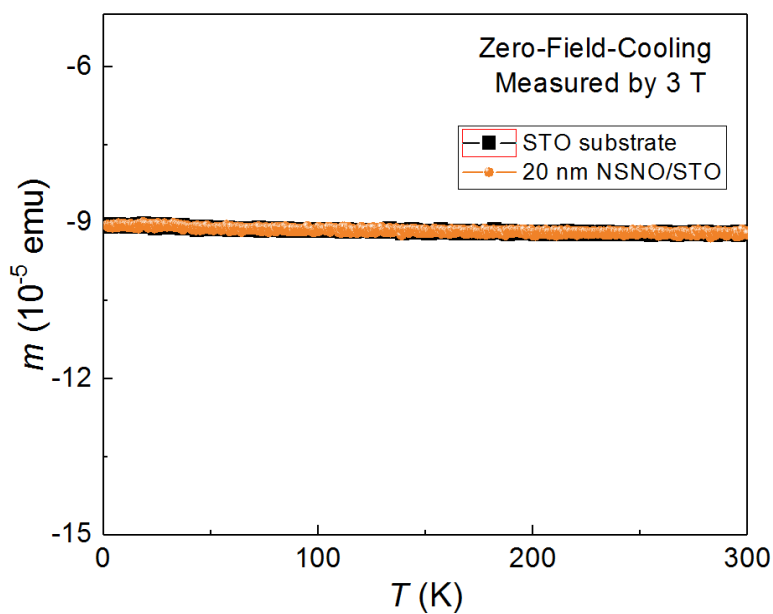

**Figure S2** Temperature-dependent total magnetic moment of an STO substrate and a 20-nm-thick NSNO/STO heterostructure ($5 \times 2.5 \times 0.5$ mm$^3$) measured by an in-plane field of 3 T under the zero-field-cooling measurement mode.

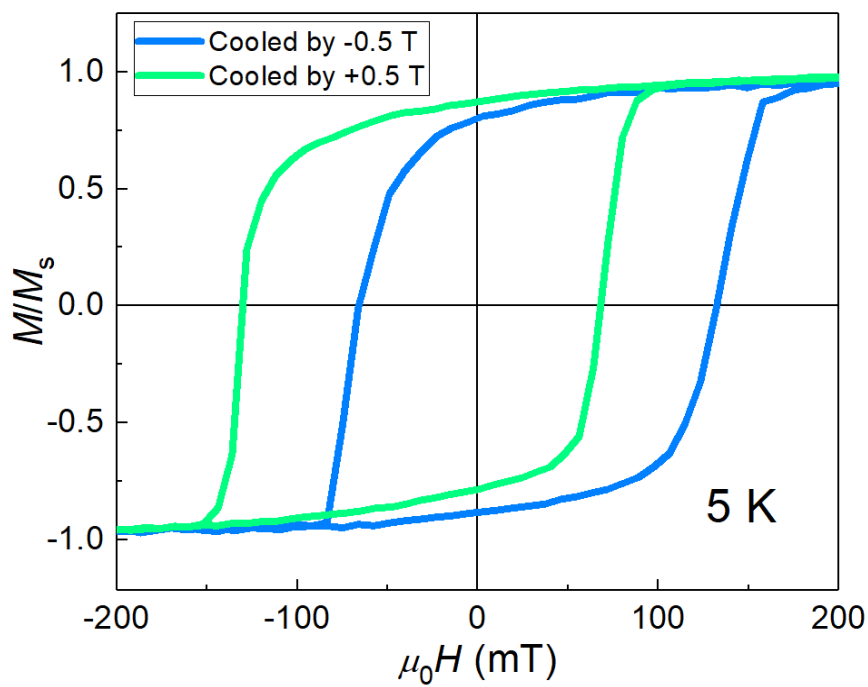

**Figure S3.** Normalized magnetization versus magnetic field for a Pt (2 nm)/CoFe (5 nm)/NSNO (20 nm)/STO sample at 5 K obtained by field cooling measurements with a cooling field of +500 and -500 mT, respectively. The cooling process was started from 300 K.

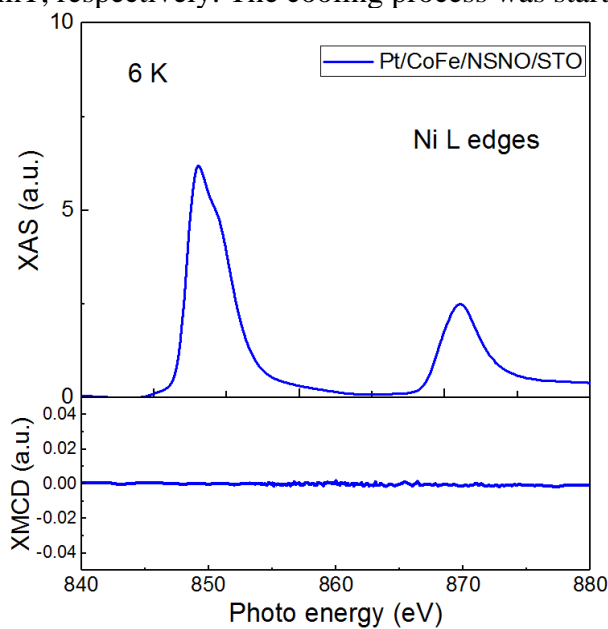

**Figure S4.** X-ray absorption spectra of Ni L edges in a Pt(2 nm)/CoFe(5 nm)/NSNO(20 nm)/STO heterostructure at 6 K and the corresponding X-ray magnetic circular dichroism signal. The spectra were collected by the total electron yield (TEY) mode with a grazing incidence and an in-plane magnetic field of 1 T. The numbers on the *y* axes of the XAS and XMCD spectra are of the same scale.

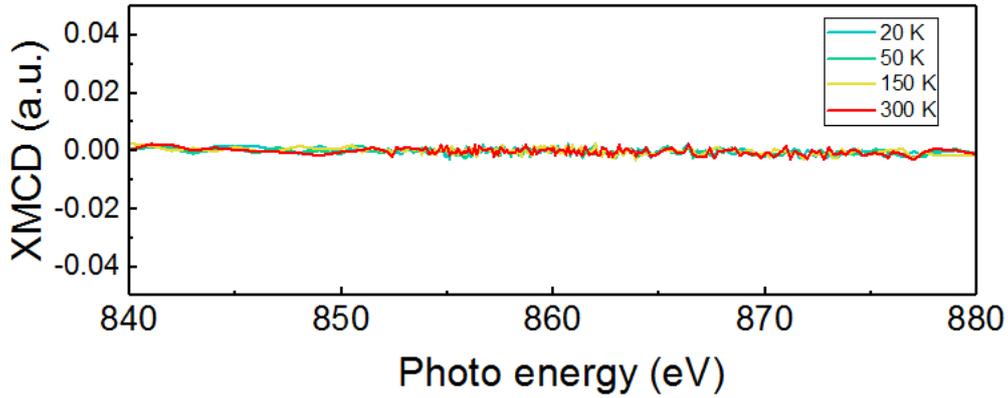

**Figure S5.** X-ray magnetic circular dichroism spectra of Ni L edges in a Pt(2 nm)/CoFe(5 nm)/NSNO(20 nm)/STO heterostructure at higher temperatures. The spectra were collected by the TEY mode with a grazing incidence and an in-plane magnetic field of 1 T. The numbers on the *y* axis is of the same scale with those in the XAS and XMCD spectra of Figure S4.

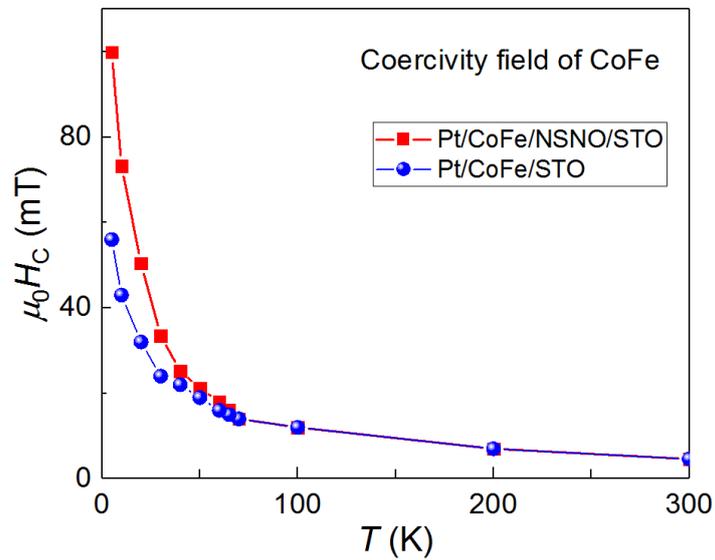

**Figure S6.** Temperature-dependent coercivity field of the CoFe films in the Pt(2 nm)/CoFe(5 nm)/STO and the Pt(2 nm)/CoFe(5 nm)/NSNO(20 nm)/STO samples.

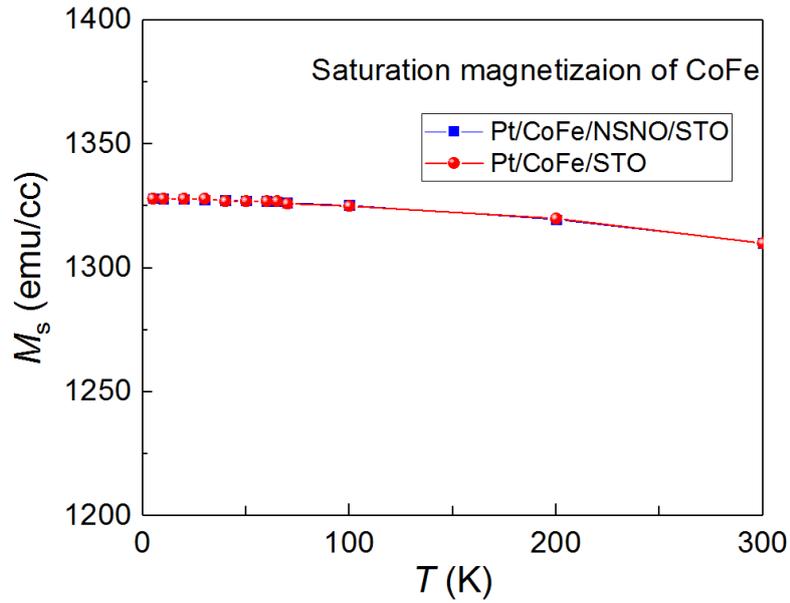

**Figure S7.** Temperature-dependent saturation magnetization of the CoFe films in the Pt(2 nm)/CoFe(5 nm)/STO and the Pt(2 nm)/CoFe(5 nm)/NSNO(20 nm)/STO samples.

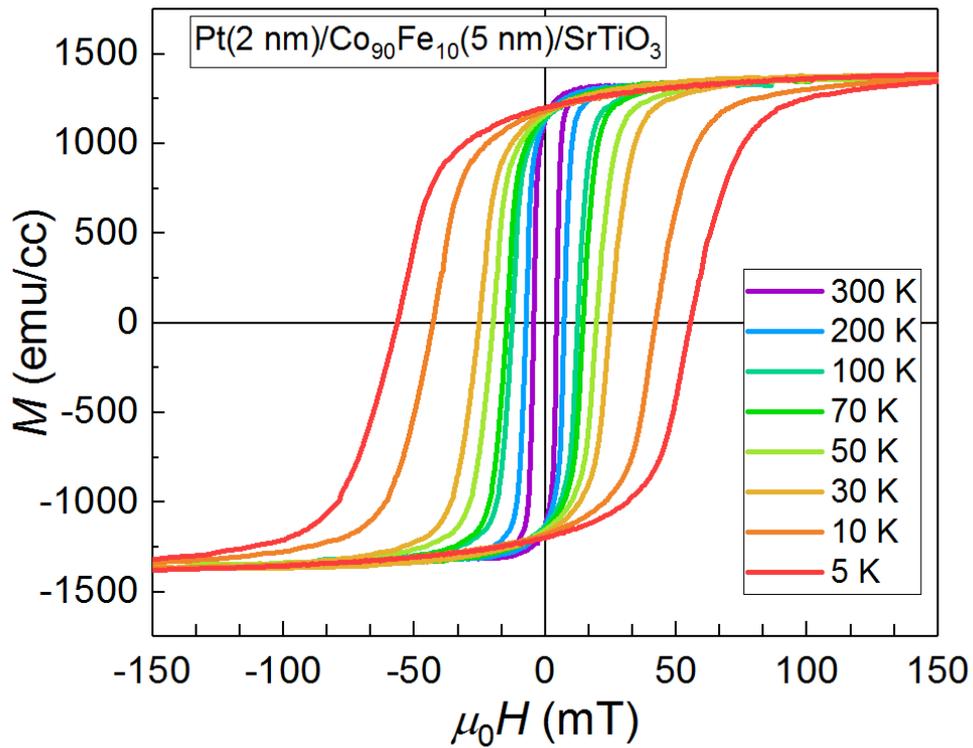

**Figure S8.** Magnetization versus magnetic field for a Pt(2 nm)/Co$_{90}$Fe$_{10}$(5 nm)/SrTiO$_3$ reference sample measured by in plane magnetic fields at different temperatures. The measurements were performed by the field cooling mode with a cooling field of +500 mT and the cooling process was started from 300 K.

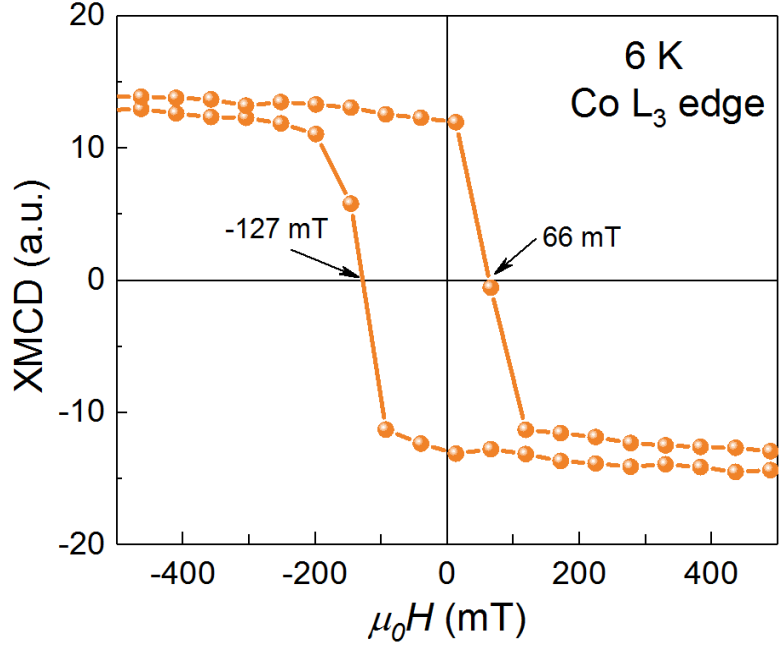

**Figure S9.** X-ray magnetic circular dichroism intensity of the Co $L_3$ edge in a Pt(2 nm)/CoFe(5 nm)/NSNO(20 nm)/STO heterostructure at 6 K. Before the measurement, the sample was cooled from 300 K under an in-plane magnetic field of 500 mT. The spectra were collected by the TEY mode with a grazing incidence. It shows an exchange bias field of ~30 mT, which is consistent with the magnetic hysteresis loop measured by SQUID shown in **Figure 3**.

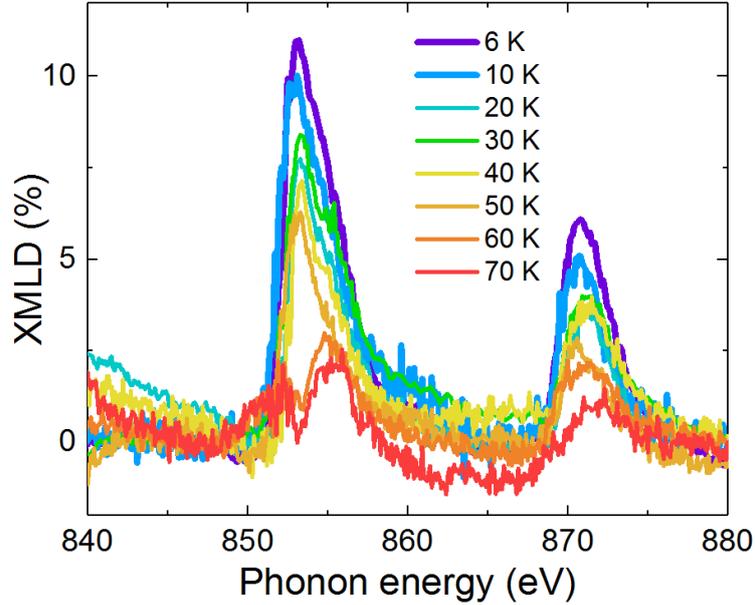

**Figure S10.** XMLD spectra of the 20-nm-thick NNSO/STO heterostructure at various temperatures obtained by XMLD (%) = $(I_{Normal}-I_{Graze})/(I_{Normal}+I_{Graze}) \times 100\%$, where $I_{Normal}$ and $I_{Graze}$ are the absorption intensity of differently incident light shown in **Figure 5** of the main manuscript.

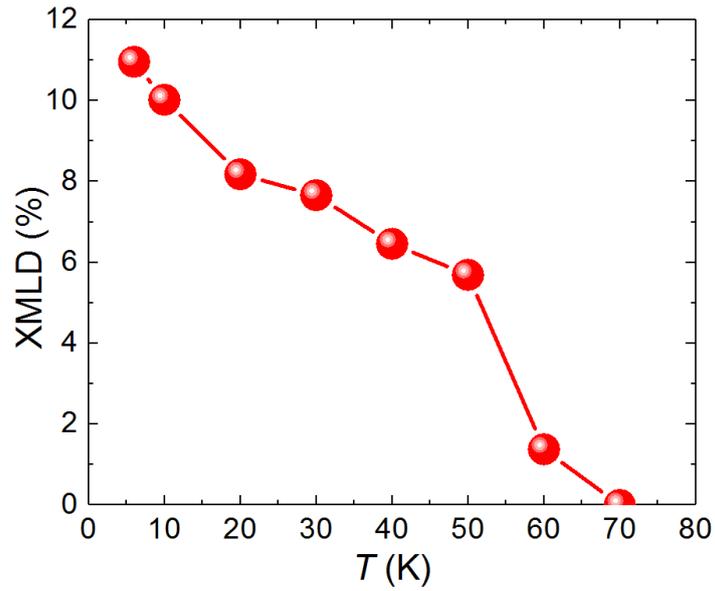

**Figure S11.** The relative X-ray magnetic linear dichroism signal for the 20-nm-thick NNSO/STO heterostructure at ~853 eV (L$_3$) at various temperatures.

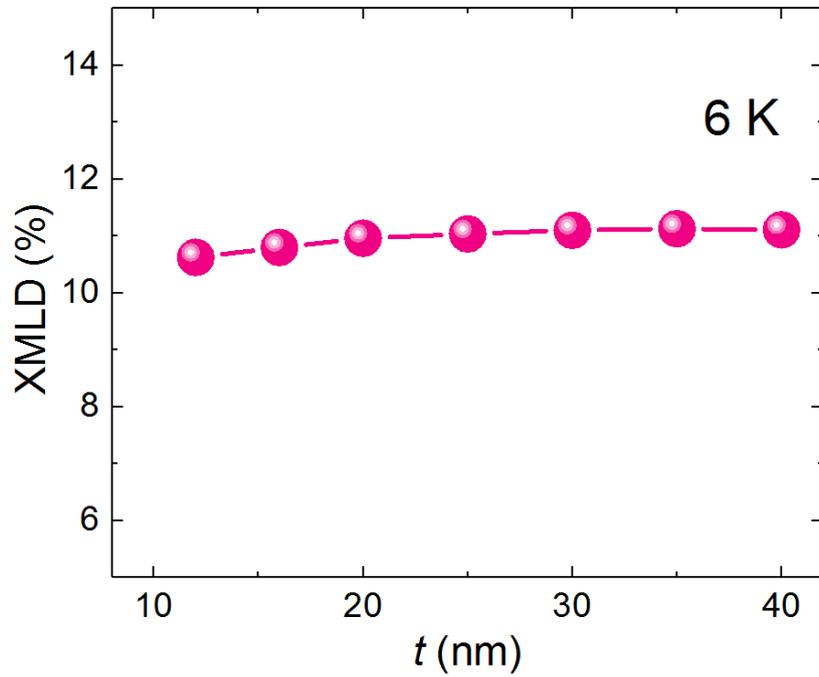

**Figure S12.** Thickness-dependent relative XMLD signal at ~853 eV (L$_3$) for NSNO films at 6 K for NSNO films thicker than 8 nm. The critical temperature for the existence of XMLD signal of 8-nm-thick films is ~5 K.

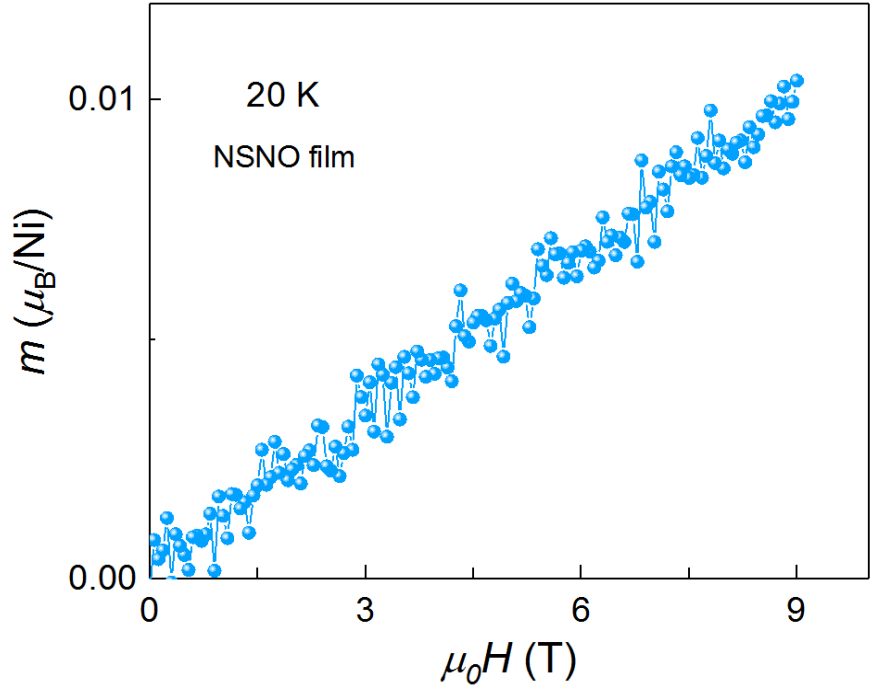

**Figure S13.** Magnetic-field dependence of the average Ni moment (assuming that all the moments of the NSNO film were carried by Ni atoms) in a 20-nm-thick NSNO film. It is obtained by performing magnetic measurements up to 9 T at 20 K for the 20-nm-thick NSNO/STO heterostructure and an STO substrate with the same dimensions and then conducting the subtraction.

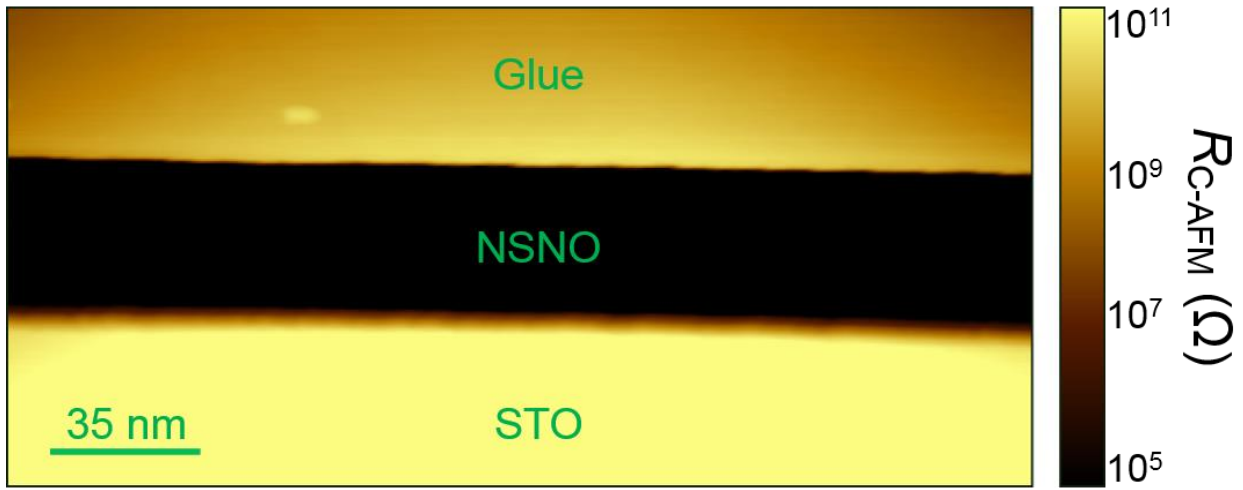

**Figure S14.** Cross-section conduction mapping of a 35-nm-thick NSNO/STO heterostructure that exhibits the highest $T_{c\ Zero}$ (~12.7 K) by a low-temperature conducting atomic force microscopy setup at 10 K. The cross-section sample was prepared by cutting the NSNO/STO sample into two pieces and pasted them with glue face to face. Subsequently, mechanical polishing with diamond-coated discs and colloidal silica were performed to achieve a fine cross-section surface.

# Note S1: Estimation of local Ni moment in an NSNO film

A schematic for the spin canting model in NSNO induced by an external magnetic field is shown below.

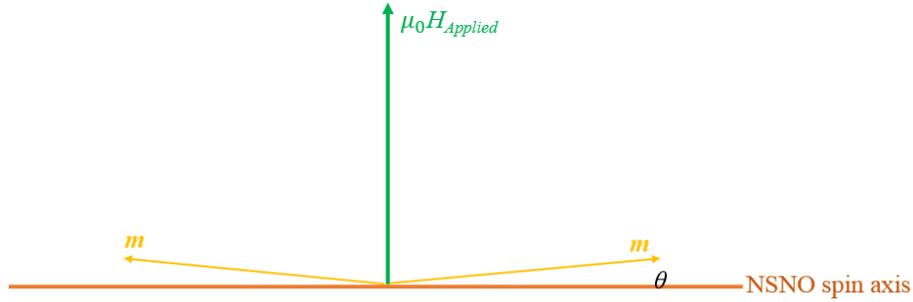

**Figure S15.** Schematic for the external magnetic field induced spin canting in a collinear antiferromagnet. *m* represents the local sublattice magnetic moment.

For a 20-nm-thick NSNO film, we could take the blocking temperature ~65 K as an estimation of the Néel temperature, which then corresponds to an antiferromagnetic exchange coupling field $\mu_0 H_E$ ~97 Tesla as the Zeeman energy of 1 Tesla $\mu_B B$ is equal to the thermal energy of 0.67 K.

As a result, under an external magnetic field of 9 T, the field-induced spin canting angle $\sin\theta = \frac{1}{2}\mu_0 H_{Applied}/\mu_0 H_E \approx 9/194$.

Accordingly, the induced moment for each Ni sublattice is $m\sin\theta \approx 9m/194 \approx 0.01~\mu_B$ (experimentally measured value for 9 T as shown in **Figure S13**). From this equation, we can infer that the upper limit of the local moment for Ni is $m \approx 0.22~\mu_B$.

# Note S2: Theoretical calculations on the surface-induced antiferromagnetic order in NSNO thin films

Our density functional theory calculations reveal that the surface of NdNiO$_2$ film could largely stabilized the antiferromagnetic order. To simulate the surface effect, we construct $\sqrt{2} \times \sqrt{2} \times 4$ supercell and adopt a large vacuum layer at two sides. The two surfaces are both constructed by Nd layer which are dealt with two cases, including with (NdO surface) or without (Nd surface) apical O atom, as shown in **Figure S16**. Two different antiferromagnetic configurations, *i.e.*, *G*-type and *C*-type antiferromagnetic orders as shown in **Figure S16**, are considered in calculations. The calculations were performed using Vienna *ab initio* simulation package (VASP) and Perdew-Burke-Ernzerhof (PBE) functional. The calculated results are shown in **Table S1**. We have considered the undoped case and 0.2-hole-doped case (to simulate 20% Sr doping case). The energy of the system with ferromagnetic order is set as a reference, and hence the energy in this table is $E_{AFM}-E_{FM}$ per Ni atom. For bulk NdNiO$_2$, we have calculated the energy of various kinds of magnetic order, and found that different magnetic order competes with each other and a strong spin fluctuation occurs as reported in previous experimental and theoretic works [Nomura *et al.*, *Phys. Rev. Research* **2**, 043144 (2020)]. Such a behavior gives a rather small negative energy of antiferromagnetic order as shown in **Table S1**. For the hole doped bulk NdNiO$_2$, the relatively large positive energy of antiferromagnetic order indicate that the antiferromagnetic order is strongly suppressed. When the surface effect is included, no matter for NdNiO$_2$ film with or without apical O atom, and undoped or hole doped cases, the large negative energy of antiferromagnetic order suggests that such order could be stabilized.

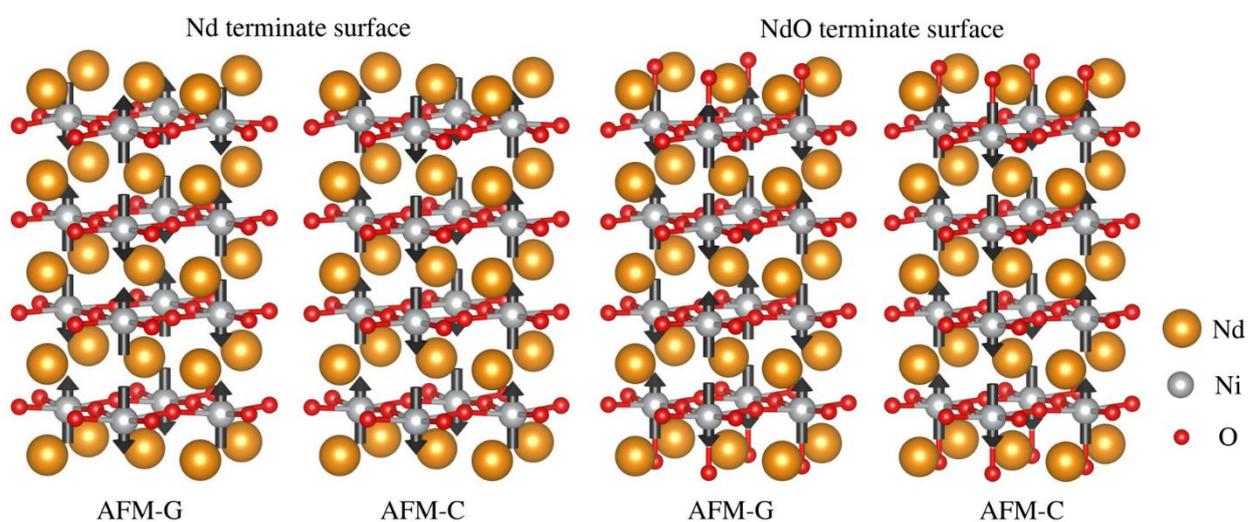

**Figure S16.** The crystal structures of NdNiO$_2$ film with or without apical O atom at surface and corresponding antiferromagnetic orders considered in calculations.

**Table S1.** The calculated energies (meV per Ni atom) of different antiferromagnetic orders of bulk NdNiO$_2$ and NdNiO$_2$ film. The undoped and 0.2-hole-doped cases are both considered.

|  | Bulk NdNiO$_2$ | | NdNiO$_2$ Thin Film | | | |
| --- | --- | --- | --- | --- | --- | --- |
|  |  |  | Nd surface | | NdO surface | |
|  | undoped | doped | undoped | doped | undoped | doped |
| AFM-*G* | -19.69 | 150.79 | -263.79 | -59.40 | -153.86 | -76.36 |
| AFM-*C* | -18.41 | 32.51 | -251.04 | -41.94 | -192.10 | -106.07 |